\documentclass[conference]{IEEEtran}
\IEEEoverridecommandlockouts
\usepackage{cite}
\usepackage{amsmath,amssymb,amsfonts}
\usepackage{algorithmic}
\usepackage{graphicx}
\usepackage{textcomp}
\usepackage{xcolor}
\def\BibTeX{{\rm B\kern-.05em{\sc i\kern-.025em b}\kern-.08em
    T\kern-.1667em\lower.7ex\hbox{E}\kern-.125emX}}
\begin{document}

\title{Smart Switching and Control of a Distributed Generator Synchronized With National Grid
\\
\thanks{Authors would like to thank DEE, PIEAS, Pakistan for funding and supporting this research work.}
}

\author{\IEEEauthorblockN{Muhammad Sarwar\text{*}, Atiqa Dawood, Maheen, Ujyara Saleem, Muhammad Abubakar}
\IEEEauthorblockA{\textit{Dept. of Electrical Engineering} \\
\textit{Pakistan Institute of Engineering \& Applied Sciences}\\
Islamabad, Pakistan \\
\text{*}msarwar@pieas.edu.pk}
}

\maketitle

\begin{abstract}
Distributed generation is widely being utilized, so the basic theme of this research is to have a hands-on experience to synchronize a Distributed Energy Resource (DER) to the Mains Grid. A control algorithm is implemented for energy supply from both sources i.e. DER and Mains Grid, according to the cost and availability of each source. When load changes, so that the synchronization does not lose. LabVIEW software is used for the implementation of desired system. Validation is done by experimenting in lab. In order to keep the system stable, the algorithm has been developed to shift the load from one bus to the other depending upon the load requirements. The project is a test bed and can be used for further experimentation and prove helpful in developing the basic understanding of Smart Switching using LabVIEW.\\
\end{abstract}


\section{Introduction}
The electricity we obtain from grid stations can be cleaner and cost efficient. The main concern is rising cost of electricity. The electricity we obtain from grid costs 31 percent more during peak hours. The other resources which can be used for this purpose are being ignored for numerous reasons. The switching time and decision of DER is important. Manual Switching is hectic and not really practical. This can be automatic if we take into account the parameters. We have considered Distributed generator as source other than grid. The peak hour cost can be reduced to a considerable amount if we shift the load to DG for those hours. For the simulation purpose LabVIEW is used that is user-friendly and provide more leniencies for further research as well.

After the arrival of electrical distribution grids, the demand of devices for the measurement of consumption of energy, distribution, price and monitoring the services increased rapidly. Previously there were tentative devices used for the measurement of consumption of energy which has now being replaced with smart grid, which is a two-way metering technology used to turn off and on the appliances according to the demand of electricity and off-peak prices. In the beginning many hardships were faced to understand the behavior and working of grid.

Switching between multiple generating units depending upon loads and resources is called smart switching. With the help of smart switching the energy industry can be made more available, reliable and efficient. During the period of transmission it will be more efficient, management and operation cost for utilities will be reduced, security will be improved and integration of customer-owner power generation systems and renewable energy systems will be improved. It also helps to save money by managing electricity use and the best times to purchase electricity is chosen. With smart switching we can switch the power generating devices which can also prevent blackouts in case of shortage of electricity.

DERs and their integration with Conventional Power Grid is an active research topic among industry and academia researchers. Current research being carried out in this field consist of integration of DGs with grid, building test-bed systems for Smart Grid research. A thorough review of smart grid technologies for the future power systems is given in \cite{sarwar2016review}. Power control in a microgrid through active synchronization is explored in \cite{cho2011active}. A smart monitoring system for renewable energy microgrid using LabVIEW is designed by N. Chinomi \textit{et al} in \cite{chinomi2017design}. Authors from reference \cite{sardar2008synchronous} implemented equations in LabVIEW to simulate behavior of a synchronous generator. Reference \cite{Navitha2018efficient} discusses efficient energy management of a microgrid through real-time simulations in LabVIEW.

Authors from \cite{aissou2015modeling} propose a method to integrate renewable energy resources to conventional grid along with energy storage. Grid synchronization studies have been performed by authors of \cite{petersson2005modeling,mohammed2010laboratory,martis2006electrical}, they discuss synchronization problems of Doubly Fed Induction Generator (DFIG) and various other connectivity issues. A LabVIEW based test-bed system has been designed by authors of \cite{mahmood2019design} where they designed and test an automatic synchronizing and protection relay for the distribution generator (DG).

LabVIEW is an obvious choice for developing control and interfacing for data acquisition and real-time control of instruments.

DERs are very suitable for meeting the monitoring and management requirements of the ever changing grid conditions. A secure and reliable infrastructure for the linkage of DER asset with intelligent control systems is required for doing this in a cost effective manner.

DERs can significantly enhance the performance of grid. Utilities can integrate DERs to cater the demand response for the reduction of peak-load. During peak hours energy capacity is very high than the usual energy requirement, DERs are used to cater this energy management in a cost effective manner, whether it includes integrating rooftop solar system or using local distribution system for reducing dependence on costly and carbon intense generation capacity.

Laboratory Virtual Instrument Engineering Workbench is a graphical programming language. LabVIEW has proven to be an invaluable tool which can decrease development time in design, research, and manufacturing. Besides this, LabVIEW is a user friendly and easy to use software even for someone with very less knowledge of LabVIEW.

\subsection{Objectives}
The objective of was to have a hand on experience in order to synchronize a Distributed Energy Resource (DER) to the Mains Grid. This is done by controlling the supply of energy from both sources (i.e. WAPDA and DER) by making a comparison between the cost and availability of sources when load changes, without loss of synchronization. Synchronization is done by developing custom auto-syncing modules in LabVIEW software. After that switching between both sources is done using 8 switch Relay.

\subsection{Organization of Paper}
The rest of paper is organized as follows:

\section{Synchronization of National Grid and Distributed Generator}

\subsection{Methodology}
First of all, literature regarding DERs and Synchronous generators has been studied. Manual Synchronization is done practically using the setup available in Power Lab. The working of PID is studied and used for automatic synchronization using LabVIEW for simulation purpose. Simpler loops are designed in LabVIEW for speed control that can be customized easily. Speed Control and Voltage control loops are designed for comparing the values from grid and DG. Before the synchronization is done all the conditions are met. At the end, load switches automatically, making the system cost efficient.

\subsection{Hardware Setup}
The hardware setup consists of the Distributed Generator, supply from WAPDA, arduino board, loads, eight channel power relay and the LabVIEW software.

\begin{figure}
    \centerline{\includegraphics[width=0.5\textwidth]{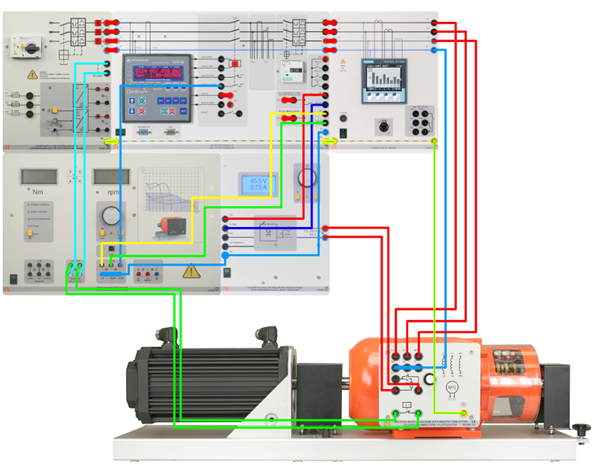}}
    \caption{Hardware Setup}
    \label{fig1}
\end{figure}

\subsection{Working}
In a synchronous generator when dc current is applied to the rotor winding a rotor magnetic field is produced. Then prime mover turns the rotor of the generator which produces a rotating magnetic field in the machine. Due to this rotating magnetic field three-phase set of voltages are produces within the stator windings of the generator. The windings on a machine are categorized as field windings and armature windings. The windings that produce the main magnetic field in a machine are termed as field windings and the windings where the main voltage is induced are called armature windings. The terms rotor windings and field windings are used interchangeably for the synchronous machines since the field windings are on the rotor. Similarly, the terms Stator windings and armature windings are used interchangeably.

\subsection{Conditions for Synchronization}
Automatic synchronization helps to switch between mains and generator without involving human interaction with system once the setup and software are set for required settings. The conditions for automatic synchronization are frequency synchronization, voltage synchronization, phase angle synchronization \cite{chapman2005electric}.

\section{Frequency Synchronization}
Synchronous machine is known as constant speed machine and also independent on load. Speed control is the most important task regarding synchronous machine operation in real time. One way is to control speed by varying frequency. It can also be achieved by voltage fed inverters or current fed inverters.

In order to make the speed of generator equal to that of the speed of grid we proposed a new loop. This loop was designed in such a way that first it compares the speed of generator and the speed of grid. The speed of generator is taken directly from the system and the speed of grid is calculated by using the frequency of grid and the number of poles, using the equation, as n = 120f/p.  According to the designed loop if the difference between the generator’s speed and grid’s speed is more than 500 rpm then an increment of 50 rpm is done to the generator speed in a continuous loop until the generator’s speed becomes equal to that of the grid’s speed. If  the difference between the generators speed and the grids speed is between 100 rpm and 500 rpm then an increment of 10 rpm is done to the generators speed at a time till the generators speed becomes equal to that of the grids speed. If the difference between the generators speed and the grids speed is between 10 rpm and 100 rpm then an increment of 2 rpm is added to the generators speed at a time till the generators speed becomes equal to that of the grids speed. The tolerance of the loop is 2 rpm that is if the difference between the generators speed and the grids speed the loop stops. 

In result of the above loop simulation the generator’s speed reaches the grid’s speed as seen from graph 5.3 when grid’s speed is 400 rpm then the generator speed also reaches 400 rpm after a few seconds.

\subsection{Interface for Generator’s RPM}
\begin{figure*}
    \centerline{\includegraphics[width=\textwidth, height=5cm]{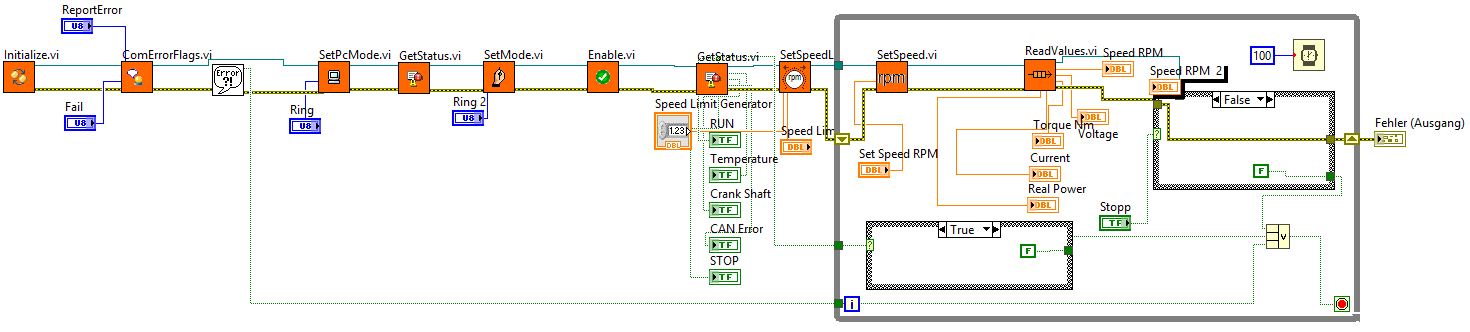}}
    \caption{VI for obtaining Generator’s RPM}
    \label{fig2}
\end{figure*}

Generator’s RPM is obtained by using a virtual instrument in LabVIEW as shown in Figure \ref{fig2}.

\subsection{Low Pass filter}
\begin{figure}
    \centerline{\includegraphics[width=0.5\textwidth, height=4cm]{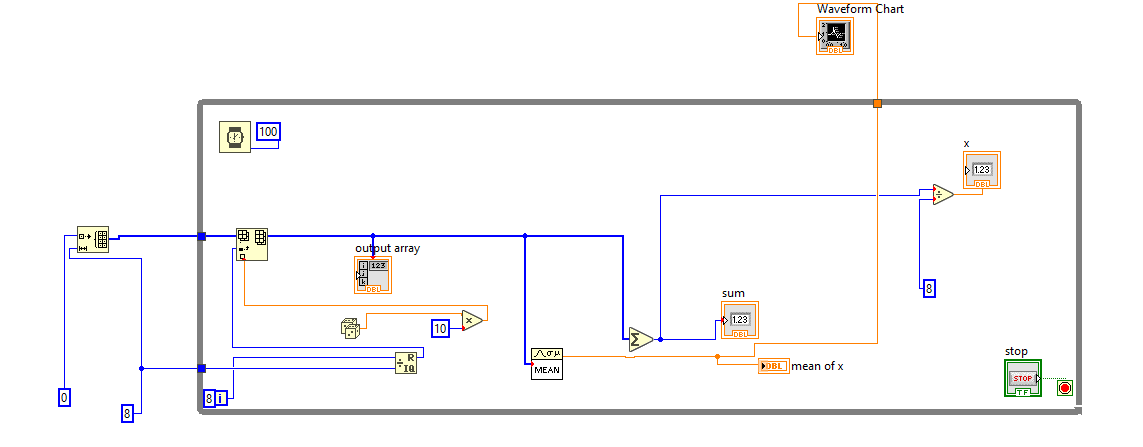}}
    \caption{Low pass filter to remove glitches}
    \label{fig3}
\end{figure}

When we work in real time and take values directly from grid, smooth value is not expected. The input is a slightly varying frequency. The variations are not very large but we need a comparatively smooth value we designed a filter that takes 8 values and average them. This average value is fed to the system making it a smooth running system while taking values from grid and it will help to synchronize faster. Initially values are stored in array after 8 values another relay reads the values and them to an adder that add up all the values. This sum is sent to a divider that divides the value with a constant 8 each time. This average value is the output and also displayed using display. Since this is in a while loop it continues to repeat until met the final condition or switched off. The VI built is shown in Figure \ref{fig3}.

\subsection{Grid’s Frequency from Power Meter}
Grid's frequency is obtained from quality meter which is then converted into speed as
\begin{equation}
N = 120\frac{f}{P}
\label{eq1}
\end{equation}

\begin{figure}
    \centerline{\includegraphics[width=0.5\textwidth]{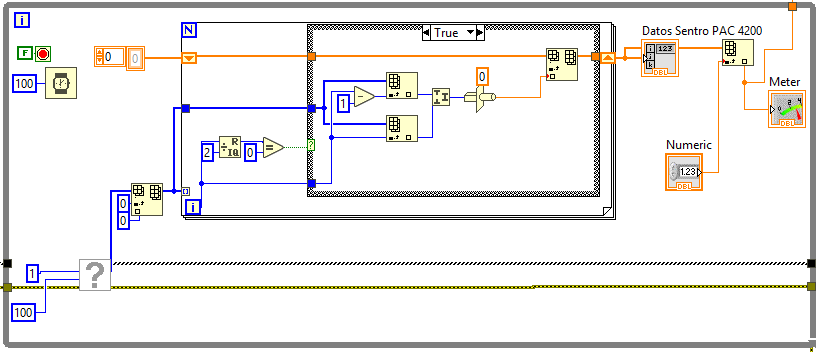}}
    \caption{VI for obtaining grid’s frequency}
    \label{gridFreq}
\end{figure}

\subsection{Speed Synchronization Loop}
For the sake of automation the task was completed by using software LabVIEW. Speed control was achieved by a built in VI with some modifications to make it useful practically. A delay (100 ms in this case) was inserted which helps it to operate smoothly with each variation. The assembly shown in picture has many units involved for speed control. 

\begin{figure}
    \centerline{\includegraphics[width=0.5\textwidth, height=4cm]{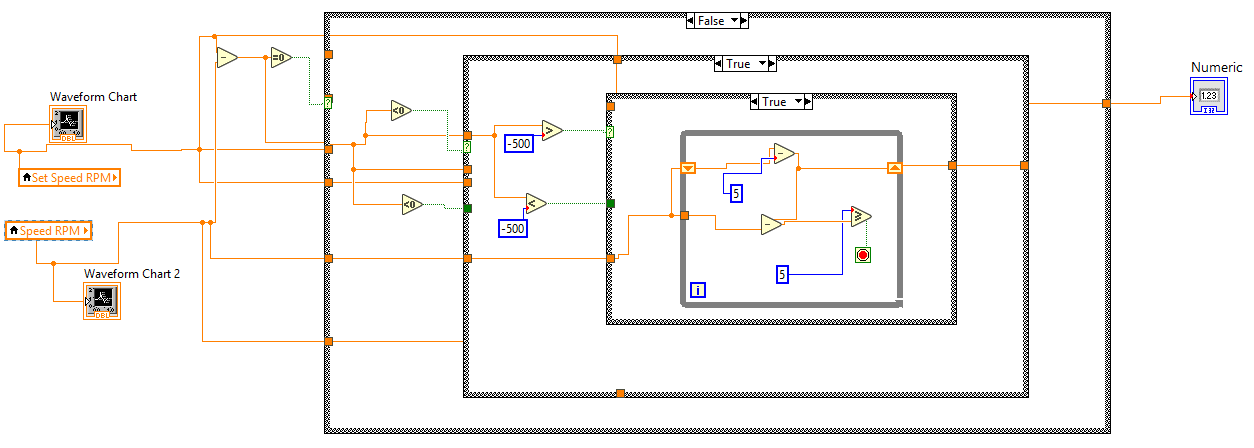}}
    \caption{Speed Synchronization Loop}
    \label{fig4}
\end{figure}

\section{Voltage Synchronization}
Our next task was the synchronization of the generators voltage and grids voltage. For this purpose we used arduino UNO and a multi functional 8 channel relay. We designed a loop for synchronizing the voltage of generator and the grid. The loop first compares the generators voltage and the grids voltage. Then according to the difference of both voltages a signal is provided to the Arduino and then to the 8 switched relay. Then a signal is provided to the ON pin of the excitation module and as a result excitation module turns ON. After that, according to the difference in the voltages of the generator and the grid a signal is sent by the loop to the UP or DOWN pin of the excitation module. If generator’s voltage is greater than excitation voltage keeps on increasing until both are same and if grid’s voltage is more than a signal is provided to DOWN button of Excitation module and excitation voltage continues to decrease until the difference between voltages is reduced to zero. The whole loop executes till the difference between the voltages drops to zero. After that a signal is sent to the OFF pin of the excitation module. Thus according to this loop the voltages of the generator and grid are synchronized till the difference between them is zero.

\subsection{Arduino UNO}
As for this task we needed to interface arduino with LabVIEW, which was best possible with arduino uno as other versions of arduino e.g. Arduino mega do not support interference with LabVIEW. The Arduino UNO is a widely used open-source microcontroller board based on the microcontroller .The board consists of sets of digital and analog input/output (I/O) pins that may be interfaced to various expansion shields and other circuits. The board consists of 14 Digital pins and 6 Analog pins. It is programmable and uses type B USB cable. An Arduino UNO can be powered by a USB cable or by an external battery it accepts voltages between 5 and 30 volts.

\subsection{Eight Channel Relay}
We used a 8 switched relay which needs an input signal of 5V for its operation. For controlling side of the relay, we need to connect our 5V power supply to the VCC and GND pins. After that we'll need to connect the IN pin to the corresponding pin of Arduino, then for the activation of relays IN pins are connected to the GND pins. On the relay side there are three main terminals of each relay. These terminals are normally known as the Normally Closed (NC) connection which is the top one, the Common (COM) connection, which is the middle one, and the Normally Open (NO) connection which is on the bottom. The relay will be connected between the NC and COM if there are no connections to the IN pin and also if the 5V power source is connected to the IN pin. The relay will connect between the NO and COM terminal if you connect the IN pin to the GND pin.

\begin{figure}
    \centerline{\includegraphics[width=0.5\textwidth, height=5.5cm]{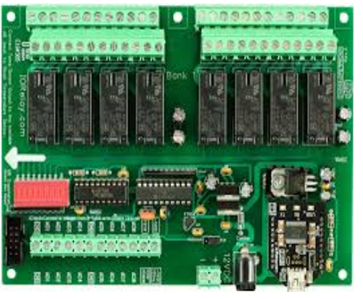}}
    \caption{8-Channel Relay for Controlling Circuit Breakers}
    \label{relay}
\end{figure}

\subsection{Loop for Voltage Synchronization}
According to the difference in the voltages of the generator and the grid a signal is sent by the loop to the UP or DOWN pin of the excitation module till the voltage of the generator and the grid is equal. 

\begin{figure}
    \centerline{\includegraphics[width=0.5\textwidth]{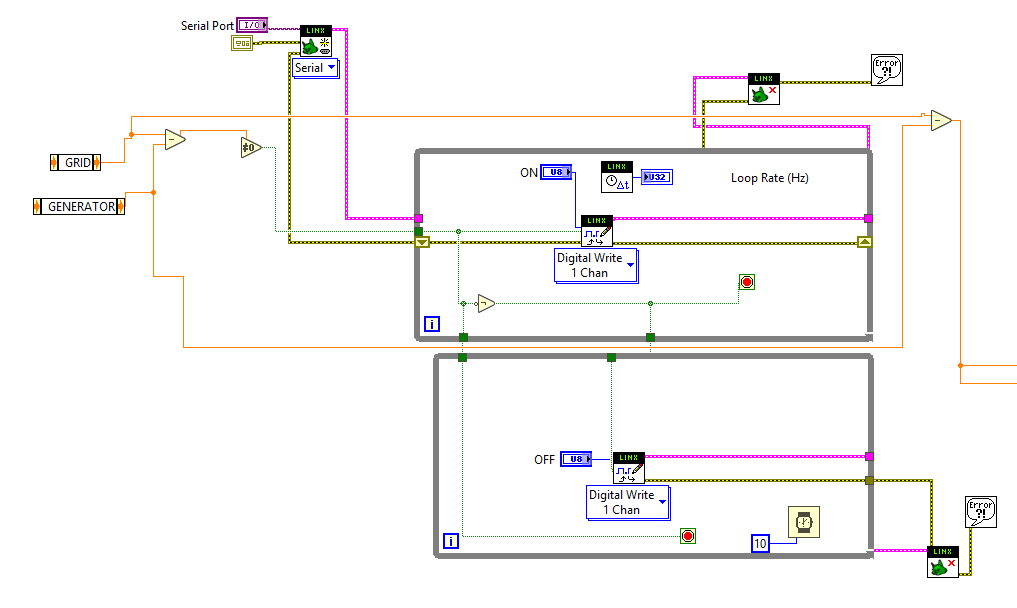}}
    \caption{Voltage Synchronization Loop - A}
    \label{fig5a}
\end{figure}

\begin{figure}
    \centerline{\includegraphics[width=0.5\textwidth]{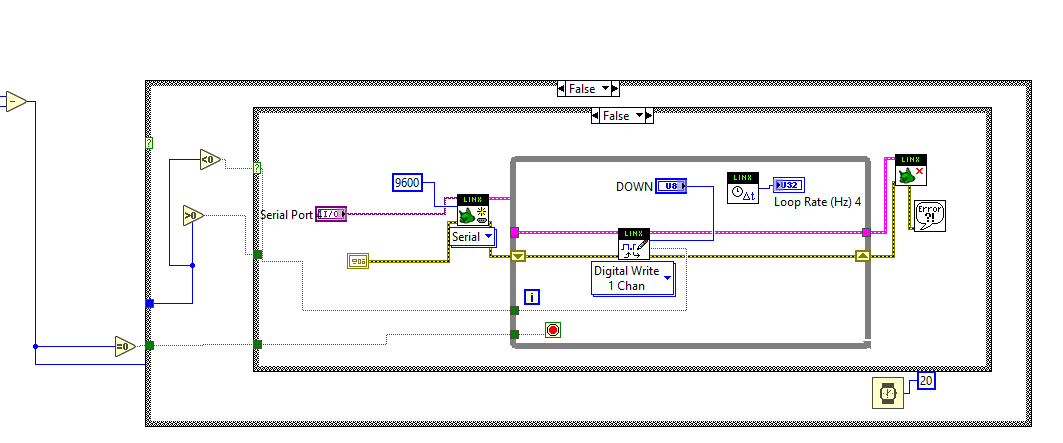}}
    \caption{Voltage Synchronization Loop - B}
    \label{fig5b}
\end{figure}

\section{Results}

\subsection{Speed Synchronization}
In result of the loop for frequency simulation the generator's speed reaches the grid's speed as seen from graph when grid's speed is 400 rpm then the generator speed also reaches 400 rpm after a few seconds.

Fig 6: Graphs showing speed of grid and generator

\subsection{Economic cost based Switching between the generator and grid}
After synchronization between grid and generator load can be shifted between them on the basis of their cost curves with respect to time. Load profile of a specific system can be shown as a graph of the variation in Load (electrical) versus time. It is bound to vary according to customer type (e.g., residential load, commercial load and industrial load) It also depend upon office timings, weather conditions and holiday seasons. Power generation companies use this data to plan how much electricity exactly they will need to make on hourly basis for the whole year. It shows nearly accurate amount of electricity that is needed for a specific region. On this basis unit cost is also decided. For peak hours (when more electricity is consumed) the cost is much higher than the cost of off-peak hours It is to discourage the use of excessive use of electricity during peak hours. From the information of Load curves and estimated costs for specific (hypothetical) system, we obtained the cost curves (Time verses the cost per unit hour) of figure 7. These curves give optimal conditions for switching making sure system runs on economically best values. We obtained the cost curves for 24 hours as in figure 8.

Time is given on x-axis and cost per unit (kwh) on y-axis. For grid the curve shows clear variations because unit cost increase during peak hours due to high usage of electricity. Here the cost on off-peak hours is 5.5 rupee/unit while for the four hours in the morning (0600hrs-1000hrs) the cost increases to 7 rupee/unit. It goes from 5.5 rupee/kWh to 9 rupee rupee/kWh. The curve shown below gives cost estimation of power obtained from generator. It has high cost initially (assuming the starting cost added up in this) and it shows a constant cost for the rest on time.

\begin{figure}
    \centerline{\includegraphics[width=0.5\textwidth]{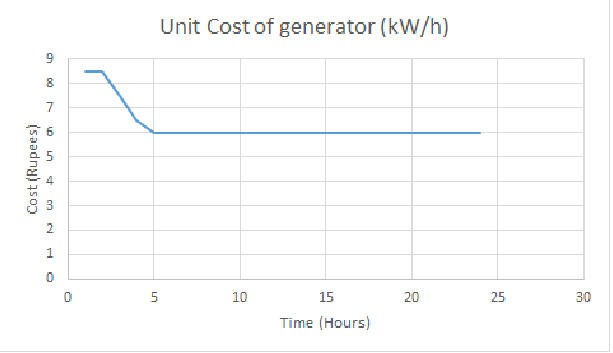}}
    \caption{Cost for system operating from Distributed Generator}
    \label{costGen}
\end{figure} 

\begin{figure}
    \centerline{\includegraphics[width=0.5\textwidth]{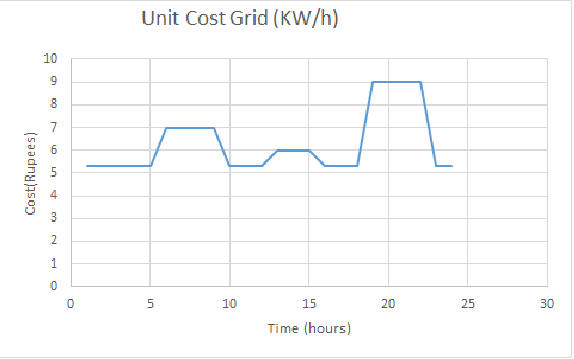}}
    \caption{Cost for system operating from grid supply}
    \label{costGrid}
\end{figure}

\begin{figure}
    \centerline{\includegraphics[width=0.5\textwidth]{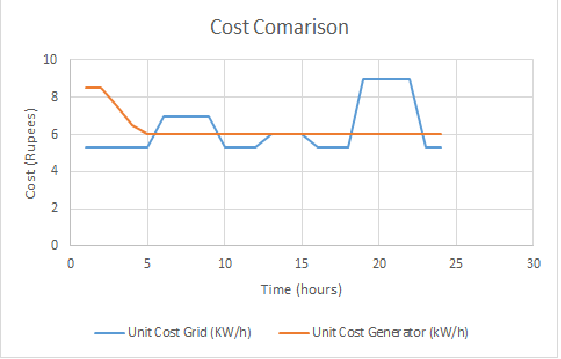}}
    \caption{Comparison of two cost curves}
    \label{costComp}
\end{figure}

Cost varies from 6rupee/kWh to 8.5 rupee/kWh. After comparing the above two cost curves, we obtained the following results. It shows the optimal cost estimation giving us idea for which time power should be obtained from generator and for how many hours switching to the grid is more economical. From the above obtained graph in figure 9,  it is clear that from 00hrs to 0600 hrs system should rely on grid supply. From 0600-1000 hrs power obtained from generator is cheaper so switching is recommended. 1000-1200 hrs operate on grid supply. From 1200-1500 hrs cost is same so switching won’t really help. From 1900-2200 hrs system will switch to generator and lastly from 2300-2400 hrs system will run on grid supply. For emergency condition, blackouts etc system can be operated on generator. If period is too long avoid the shutdown of generator, instead leave it on power savor mode to avoid the start up cost.

Power Switch module is used for switching between Synchronous generator and grid power supply after checking the necessary conditions. A loop was designed to check either the voltage and frequency is same or not. For switching it is necessary to achieve same frequency, for this purpose speed control loop was designed since we can’t change the frequency of grid supply the speed of synchronous machine is changed. Likewise same voltage is also required. After these conditions are met, we are ready to switch from grid to synchronous machine. For this purpose power switch module is used.

Power switch module consists of on and off buttons as shown in picture. This module is also connected to Arduino.
Specified pin of Arduino turns on if all conditions are true. This pin is connected to channel 5 of multi-functional relay which provided 26v at the output after receiving ON signal from Arduino these 26V are connected to ON button of Power Switch Module. So, If conditions are met this switch indicates ON and load is connected to synchronous machine and it starts acting as generator and supplies power to load. 
 
\begin{figure}
    \centerline{\includegraphics[width=0.5\textwidth]{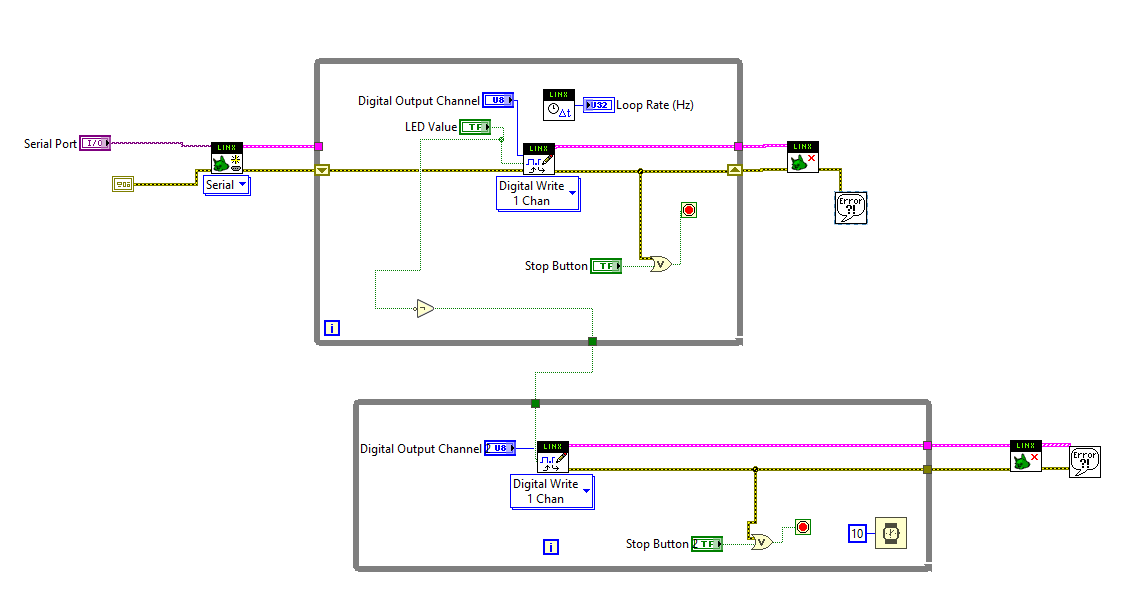}}
    \caption{Control of Circuit Breakers through Arduino}
    \label{arduioSwitch}
\end{figure}


\section{Conclusion}

We designed a test bed system which is useful in power generation, operation and control of a Distributed Generator in grid connected and isolated mode. First we made a simulation loop for synchronization of grids frequency and generators speed. After frequency synchronization grids and generators voltage was synchronized.
The synchronization is done by developing custom auto-syncing modules in LabVIEW software. After that switching between both sources is done using 8 switch Relay.

\begin{itemize}
    \item The designed system can be extended to emulate a lab-based micro grid.
    \item Control algorithms that are designed to control real and reactive power of the DG can be modified to emulate behavior of a real power plant.
    \item The excitation system control can be enhanced by designing a simulated Power System Stabilizer.
    \item This SMIB (Single Machine Infinite Bus) system can be extended to 2-area or 3-area power system to test and design PSS algorithms, power flow control etc.
\end{itemize}

\bibliographystyle{IEEEtran}

\bibliography{references}





\end{document}